\documentclass[a4paper,12pt]{article}
\usepackage{graphicx}
\usepackage{amsmath}

\usepackage{color}
\usepackage{verbatim}
\begin{document}

\newpage
\title{Multifractal analysis of the pore space of real and simulated sedimentary rocks}
\author{Abhra Giri$^{1,2}$, Sujata Tarafdar$^2$ and Philippe Gouze$^3$, and Tapati Dutta$^{1*}$}

\maketitle
\noindent
$^1$Physics Department, St. Xavier's College, Kolkata 700016, India\\
$^2$Condensed Matter Physics Research Centre, Physics Department, Jadavpur University, Kolkata 700032, India\\
$^3$Geosciences, Universite de Montpellier 2, CNRS, Montpellier, France\\
\noindent
${^*}$ Corresponding author: Email: tapati$\_$mithu@yahoo.com\\ Phone:+919330802208, Fax No. 91-033-2287-9966
\newpage
\noindent{\bf Abstract}\\ 

It is well known that sedimentary rocks having same porosity can have very different pore size distribution.  The pore distribution determines many characteristics of the rock among which, its transport property
is often the most useful. Multifractal analysis is a powerful tool that is increasingly used to 
characterize the pore space. In this study we have done multifractal analysis of pore distribution on  
sedimentary rocks simulated using the Relaxed Bidisperse Ballistic Model (RBBDM).
The RBBDM can generate a $3-D$ structure of sedimentary rocks of variable porosity by tuning the fraction $p$ of particles of
two different sizes. We have also done multifractal analysis on two samples of real sedimentary rock to compare with 
the simulation studies. One sample, an
oolitic limestone is of high porosity ($~ 40\%$)while the other is a reefal  carbonate of low porosity around $7\%$. $2-D$ sections of X-ray micro-tomographs of the real rocks were stacked sequentially 
to reconstruct the real rock specimens. Both samples show a multifractal character, but we show that RBBDM gives a very realistic representation of a typical high porosity sedimentary rock.\\


\vskip 0.5cm

\section{Introduction}
Sedimentary rocks are often the storehouses of natural oil and gases whose extraction depend on the permeability of these fluids through them. The transport properties of sedimentary rocks depend not only on the porosity of the rocks but more importantly on the pore size distribution (PSD) and their connectivity. The pore space can be a continuum of of pores with extremely varying pore sizes ranging over a scale of $10^6$,  besides being extremely complex and heterogeneous and often self-similar.

Fractal and multifractal analysis are increasingly used to study complex heterogeneous systems which show 
self-similarity on several length scales. They have the ability to provide an accurate representation of the heterogeneous 
pore geometry and address the relationship 
between porosity and a range of physical processes happening in a porous medium like transport of water in soils, extraction of oil and natural gases and $CO_2$ sequestration in sedimentary rock.   Multifractal analysis has been done  using fractal models (Rieu and Sposito, 1991), image analysis of two-dimensional sections of soil blocks  (Tarquis et al.,2003; Dathe et al.,2006; Grau et al.,2006), analysis of three-dimensional pore systems reconstructed by 
computer tomography (Tarquis et al.,2007), mercury intrusion porosimetry (Vidal Vazquez et al.,2008) and nitrogen absorption 
isotherms (Paz Ferreiro et al., 2009).

In this work, the authors use the Relaxed Bidisperse Ballistic Deposition (RBBDM) to simulate a three dimensional porous rock structure of varying porosity and pore distribution.  In our efforts (Giri et al., 2012a, 2012b) to probe the geometry of the microstructure of the pore clusters produced by the $RBBDM$ at different porosities,
  the authors had noticed that the simulated structure had a fractal nature over different length scales. The power law exponent had different values over different length scales which hinted that the pore space might have a multifractal nature. This was further strengthened by diffusion studies in connected pore clusters. For the entire range of porosities 
  studied, diffusion was found to be anamolous with different values of diffusion exponent over different length scales.  Real rock samples were studied for comparison of simulation results, and similar signature of multifractal nature was found there! 
We shall therefore investigate whether our simulated porous structure generated with the $RBBDM$ is indeed a multifractal 
  in its pore distribution.  We shall bring out the differences in the PSD of the simulated structure at different porosities through 
  a study of their multifractal spectral dimensions. Finally we shall compare our results with similar studies done on real limestone and carbonate
  rock samples.

The details of $RBBDM$ have been discussed in earlier works by the authors (Sadhukhan et al. 2007a, 2007b, 2008, 2012) in the 
study of various transport properties like permeability and conductivity through sedimentary rocks. A brief outline of the model will 
be given here for the sake of completeness.
The basic algorithm is to deposit particles of two different sizes
ballistically. In 3-D ($2 + 1$ model), we drop square $1 \times 1\times 1$ and elongated
$2 \times 1\times 1$ ‘grains’ on a square substrate. It is well known that natural sand grains are angular and elongated (Pettijohn, 1984) , so the aspect ratio $2$ is realistic.
 The cubic grains are
chosen with a probability $p$ and elongated grains with probability
$(1 - p)$. The presence of the longer grains leads to gaps in the
structure. The porosity $\phi$, defined as the vacant fraction of the total
volume, depends on the value of $p$. For $p=1$, a compact structure is produced.  As $p$ is decreased, isolated pore 'clusters' start appearing and the porosity increases. For a specific value of $p$, the $threshold$ value, a structure spanning cluster is generated. However this cannot be called a 'percolation threshold' 
as in the case of \textit{random} percolation.  In this respect the $RBBDM$ is different from random percolation problem (Stauffer and Aharony, 1994). The $RBBDM$ being a modification of the Random Deposition Model, for $p=1$, the surface width keeps increasing with height. In fact in the limit of infinite height, 
at least one narrow structure spanning pore cluster is always present. When $p$ is gradually decreased 
to below $1$, larger grains are introduced and the grains settle on the structure following the Ballistic Deposition Model. 
The introduction of even an infinitesimal quantity of the larger grain, can close a deep surface trench creating an elongated 
pore cluster. Obviously these pore clusters are longer near the surface of the structure than at the bottom. 
The presence of a single large grain sitting atop a long pore cluster, introduces correlation between adjacent columns (Karmakar et al.,2005). As the 
fraction of large grains increase, the correlation spreads through the system. So a substrate of 
sufficient height needs to be generated before the porosity value can stabilize. 
 Our model is different from the random percolation problem as even an infinitesimally small fraction of larger grains introduces correlation between columns, thus 
robbing the system of its randomness.

\begin{figure}[!h]
\begin{center}
\includegraphics[scale=0.6]{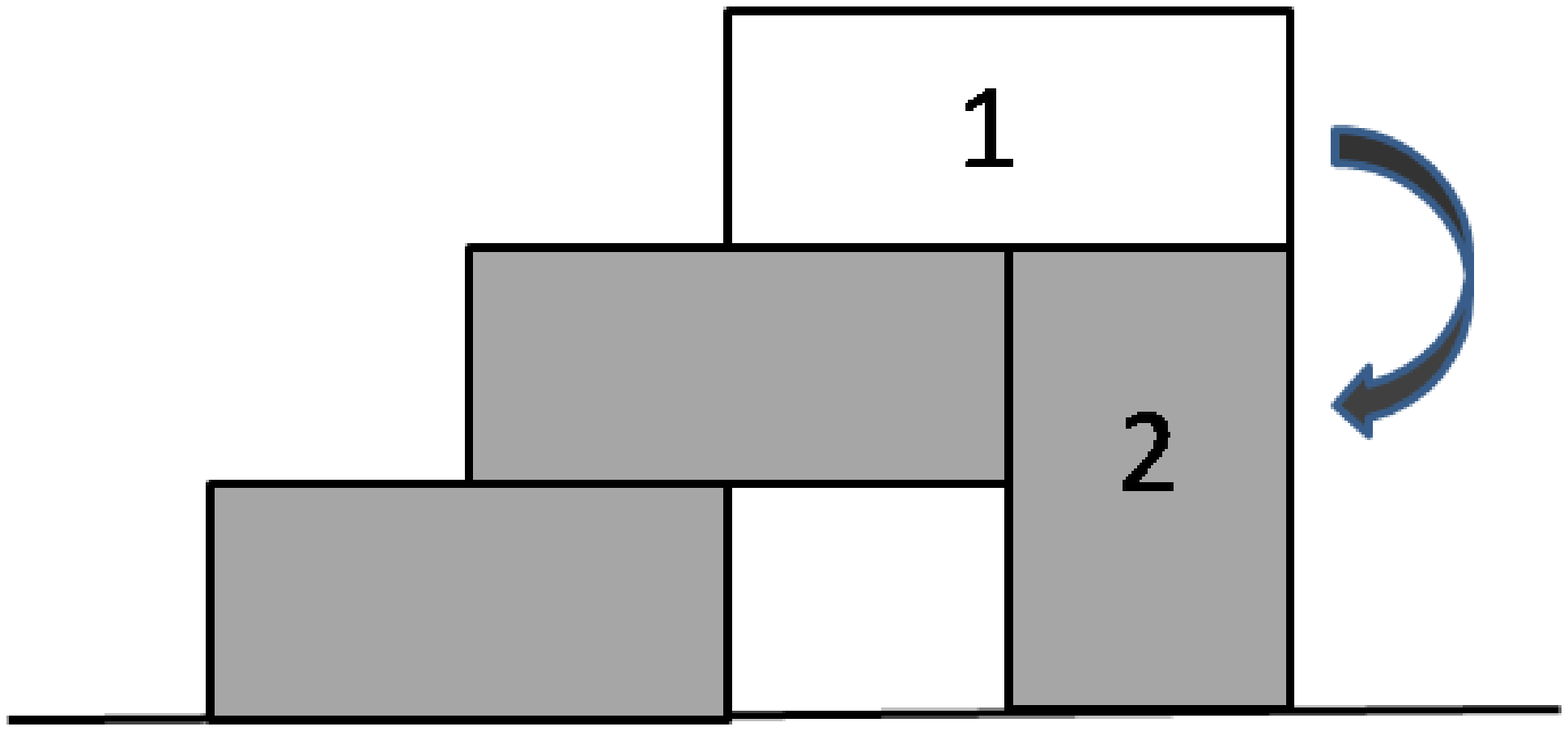}
\caption{Toppling rule of the larger grains - when a larger grain develops a two-step overhang, 
marked $1$ in the figure, with at least two vacant sites immediately below the overhang, it topples over in the direction 
indicated by the arrow to assume a more stable state, marked $2$.} 
\label{fig1}
\end{center}
\end{figure}

 The $RBBDM$ has the potential of generating a structure with a connected rock  phase that is needed for any stable structure, and a \textit{tunable porosity}. As the fraction of longer grains is increased, 
unstable overhangs can develop. If a larger particle settles on a smaller particle, a one-step
 overhang is created. 
 If a second larger particle settles midway on the previous large particle, a two-step overhang is created if there is no 
 supporting particle immediately below the protrusion of the second overhang.  This two-step overhang is not stable and the second large particle topples over if possible, according to the rule scheme as shown in fig.(\ref{fig1}). This leads to 
to compaction. 
 In their earlier works (Manna et al.2002; Dutta and Tarafdar, 2003), the authors have shown  that the sample attains a constant porosity only after a sufficient number of grains (depending on
sample size) have been deposited to overcome substrate effects. Here, a $Lx \times Ly \times Lz$ size sample was generated, from which a
$Lx \times Lx \times Lx$ sample was selected after the porosity had stabilized to within $0.001$ percent. The selected sample was chosen from below the deepest trough at the surface to 
eliminate surface effects. All simulation
was carried out on this sample. To check for finite size effects, we carried out our studies for $Lx=32,64,128,256$ for which 
$Lz=1000,2000,4000,7000$ respectively. The results reported in this work did not show any finite size dependence. All results on simulation are reported for $256\times 256 \times 256$. 

\begin{figure}[!h]
\begin{center}
\includegraphics[scale=0.45]{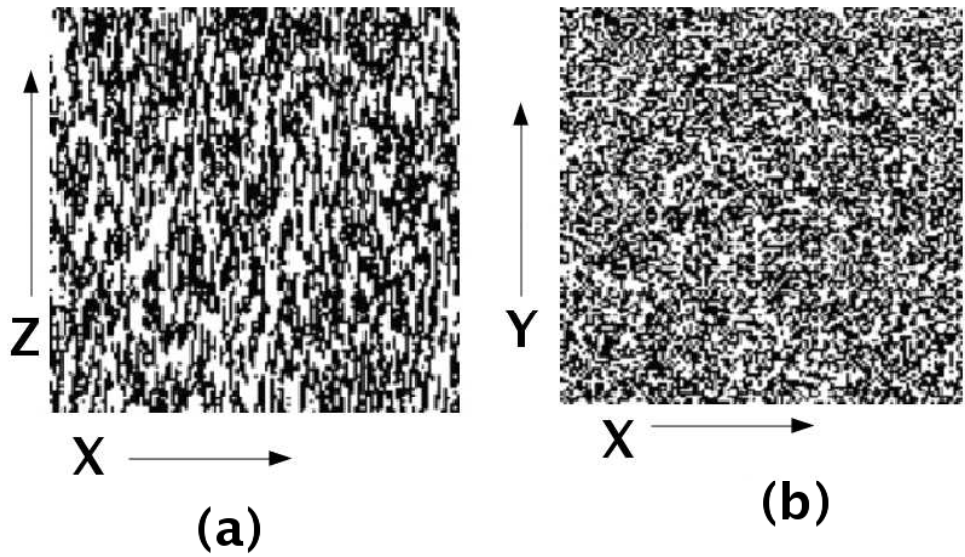}
\caption{(a) shows $x-z$ section of simulated structure for $\phi_{max}=0.45$, i.e. high porosity. (b) $x-y$
section at same porosity. Structure looks more isotropic. The white indicate pore clusters.} \label{fig2}
\end{center}
\end{figure}

\begin{figure}[!h]
\begin{center}
\includegraphics[scale=0.4]{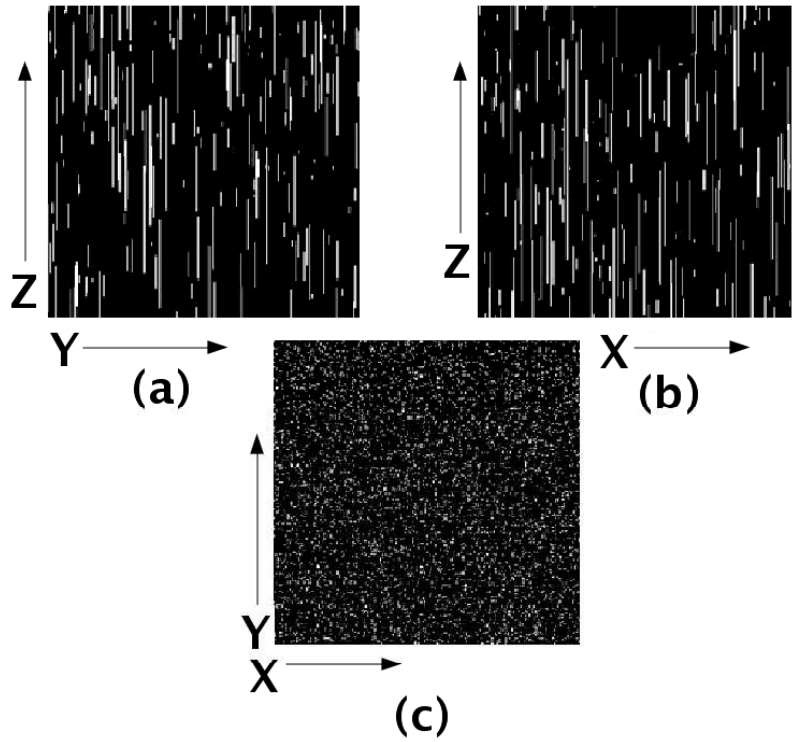}
\caption{(a) and (b) show $x-z$  and $y-z$ section of simulated structure for $\phi= 0.072 $, i.e. low porosity, matching 
the porosity of the real rock. z-axis indicates the vertical direction. The white indicate pore clusters.  (c) $x-y$
section at same porosity. Anisotropy in pore cluster structure is quite pronounced.} \label{fig3}
\end{center}
\end{figure}

 Figs.(\ref{fig2}a) and (\ref{fig2}b)
show  vertical sections,
 (x-z) and (y-z) planes of the generated sample at maximum porosity  $\phi_{max}=0.42$.
  Fig.(\ref{fig2}c) 
 shows a horizontal
section (x-y plane) of the sample at the same porosity value. The anisotropy in the 
pore geometry is clearly visible. 
The pore clusters have an elongated and interconnected appearance along the z-direction while
the distribution of pores along the horizontal plane is quite homogeneous.  As the fraction of larger grains is decreased, the porosity of the sample decreases and
 the pore distribution becomes more anisotropic nature.
  Fig.(\ref{fig3}a) and fig.(\ref{fig3}b)
  show the vertical,(x-z) and horizontal sections, (x-y)),
   respectively of the
sample at $\phi=0.073$, a very low porosity.
The elongated isolated pore clusters are prominent in the direction of assembly of the grains, whereas the pores remain homogeneously distributed in the (x-y)plane.

To compare our simulation results with real rock samples, X-ray tomography micrographs of $2-D$ sections of two real sedimentary rock samples obtained from an
oolitic limestone (pure calcite) from the Mondeville formation of Middle Jurassic age (Paris
Basin, France), and a reefal carbonate from the Majorca Island, Spain, have been used. The oolitic limestone is composed of recrystallized ooliths with a mean diameter of less
than a few hundred $\mu$m. Each pixel of both the micrographs corresponds to $5.06$ micron.   
For every real rock sample studied, each micrograph section was converted to a binary file form such that $0$ corresponded to a pore site and $1$ 
 corresponded to a rock site. The binary file was then converted to a grey-scaled picture, as shown in 
 fig.(\ref{fig4}), 
 using MATLAB. 
An array of $1000$ consecutive sections were put together precisely to reconstruct the binary file form of the 
real three dimensional rock structures. In each of the two samples, the real structure chosen was $1000 \times 1000 \times 1000$ in size, and 
all study on real rock was carried out on this structure. The  $2-D$ sections of oolitic limestone cut in the direction of assembly (growth) from the reconstructed $3-D$ structure are shown in 
Figs.(\ref{fig4}a),(x-z)plane, and (\ref{fig4}b),(y-z) plane. Comparison 
between 
figs.(\ref{fig4}a) and (\ref{fig4}b)
 shows the pore distribution to be isotropic. Fig.(\ref{fig4}c) shows a $2-D$ 
 section of the bedding plane of the same rock structure. There seems to be slight anisotropy in the pore distribution in the bedding plane and the direction of growth. Similar $2-D$ sections of the reefal carbonate rock sample have been cut along the direction of assembly of the reconstructed $3-D$ sample and shown in figs.(\ref{fig4}d and \ref{fig4}e). Fig.(\ref{fig4}f) 
 shows a $2-D$ section of the same sample cut along the bedding plane. Unlike the previous sample, any anisotropy that may be 
 present in the pore size distribution along the two directions, is not easily discernible.

  In the following section we shall briefly summarize the multifractal concepts and estimation techniques. We shall follow up with our analysis and results on the $RBBDM$. This is followed by our studies on real rock structures where we compare our findings 
  with the results of simulation. Finally we conclude with a discussion on the importance of such analysis to study pore distribution 
  in sedimentary rocks.

\section{Multifractal Concepts}
Highly inhomogeneous systems which do not obey self-similar scaling law with a single exponent, may actually consist of several intertwined 
fractal sets with a spectrum of fractal dimensions. These systems are said to be \textbf{multifractal}. Such systems have a complex distribution which arises from peculiarities of their generation. These are not a simple collection of fractal systems, rather one may say that these constitute a distribution of several fractal subsets on a geometrical support. Each of these subsets is characterized by its singularity strength and fractal dimension.

Usually multifractality arises from a spatial distribution of points, each having a \textit{strength} or \textit{weight} associated with it. The weights may also have a non-trivial distribution. A growing diffusion limited aggregate (DLA) 
with weights proportional to the growth probability assigned to each site, is an example of such a multifractal.
However, a system with equal weights assigned to each point, may form a geometrical multifractal. The distribution of pore clusters 
in a sedimentary rock belongs to this class.

The multifractal system has local fractal dimension $\alpha(x)$, which may be determined by applying the 'sandbox' method 
at different points $x$ on the system.
Different regions distributed over the system may have the same local dimension $\alpha(x)$. If we collect the $x$s with the 
same $\alpha(x)$s, we shall have identified one fractal subset of the multifractal. The scaling exponent for this subset 
has a value, say $f(\alpha)$. A plot of $f(\alpha)$ versus $\alpha$ gives a typical multifractal spectrum. For a monofractal, 
$\alpha$ would be the same everywhere and the $f(\alpha)$ versus $\alpha$ curve would reduce to a point.

Practically it is however not convenient to determine local fractal dimensions $\alpha(x)$ and therefore $f(\alpha)$. What is 
done in practice is determine the $q^{th}$ moments of distribution of points on the system, with $q$ varying from $-\infty$ 
to $+\infty$. The \textit{mass exponent} $\tau_q$ corresponding to the measure of moment $q$ has a one to 
one correspondence to $\alpha(x)$ through its derivative, and through a Legendre transformation to $f(\alpha)$. The details of 
this correspondence is discussed in the following section.

\section{Multifractal Exponents}
Multifractal analysis involves the estimation of three functions: mass exponent $\tau_{q}$, singularity strength 
(or local scaling index) $\alpha_{q}$, and multifractal or singularity spectrum $f(\alpha)$.

 Multifractal 
analysis requires that the chosen system of size $L^3$ be divided into a set of different boxes of equal 
size $\epsilon$. A common choice is to consider dyadic scaling down, i.e., successive partitions of the system in $k$ 
stages ($k=1,2,3....$) that generate a number of cells $N(\epsilon)= 2^{k}$ of characteristic length $\epsilon = L\times2^{-k}$.
 The boxes  $N(\epsilon)$ cover the entire system and each such box is labelled $i$. 
 
 The probability mass function $\mu_{i}(\epsilon)$ describing the portion of the measure contained in the $i^{th}$ box 
 of size $\epsilon$ is given by 
\begin{equation}
\mu_{i}(\epsilon) = \frac{m_i}{m_t}
\label{prob} 
\end{equation} 
where $m_i$ is the number of pore sites in the $i^{th}$ box and ${m_t}$ is the total number of pore sites in the entire system. 
A pore site refers to a pixel that is vacant. The measure of the $q^{th}$ 
 moment in the $i^{th}$ box of size $\epsilon$ is termed $\mu_{i}^{q}$. $\mu_{i}(\epsilon)$ is the total number of pore pixels inside the  $i^{th}$ box of size $\epsilon$.
Here $q$ can vary from $+\infty$ to $-\infty$.

The partition function $\chi(q,\epsilon)$  for different moments $q$ is estimated 
from $\mu_{i}^{q}(\epsilon)$ values as
\begin{equation}
\chi(q,\epsilon) = \displaystyle{\sum_{i=1}^{N(i)}\mu_{i}^{q}(\epsilon)}
\label{chi} 
\end{equation}
The parameter $q$ describes the moment of the measure. The box size $\epsilon$ may be considered as a filter so that by changing $\epsilon$ one may explore the sample at different scales. So the partition function $\chi(q,\epsilon)$ contains information at different 
scales and moments. The sum in the numerator is dominated by the highest value of $\mu_i$ for $q>0$ and the lowest 
value of $\mu_i$ for $q<0$.

The measure of the $q^{th}$ moment of the mass distribution of the system is defined as 
\begin{equation}
M(q,\epsilon) = \displaystyle{\sum_{i=1}^{N}\mu_{i}^{q}\epsilon^d} = N(q,\epsilon)\epsilon^{d}
\label{measure} 
\end{equation}
where 
\begin{equation}
N(q,\epsilon)=\sum_{i=1}^{N}\mu_{i}^{q} \sim \epsilon^{-\tau(q)}
\label{N} 
\end{equation}
If $\displaystyle{\sum_{i=1}^{N}\mu_{i}^{q}(\epsilon)}$ in the limit $\epsilon \rightarrow 0$
crosses over from $0$ to $\infty$ as $d$ changes
from a value less than $\tau(q)$ to a value greater than $\tau(q)$, then the measure has a
mass exponent
\begin{equation}
d = \tau(q)
\label{tau} 
\end{equation}
The measure is characterized by a whole sequence of exponents $\tau(q)$
that controls how the moments of probability ${\mu_{i}}$ scale with 
$\epsilon$.
For multifractally distributed measures, the partition function  $\chi(q,\epsilon)$ scales with $\epsilon$ as 
\begin{equation}
\chi (q,\epsilon)\propto \epsilon^{-\tau(q)}
\label{tau} 
\end{equation}

 The probability mass function $\mu_{i}(q,\epsilon)$ also scales with $\epsilon$ as 
\begin{equation}
\mu_{i}(q,\epsilon) = \epsilon^{\alpha_{i}}
\label{alpha} 
\end{equation}
where $\alpha_{i}$ is the $\textit{H\"{o}lder}$ exponent or 'singularity exponent' or 'crowding index' of $\mu$ peculiar to each $i^{th}$ box. Greater 
the value of $\textit{H\"{o}lder}$ exponent,  the smaller is the concentration, and  vice versa. Singularity exponents of multifractal distributions show a great 
variability within an interval ($\alpha_{max},\alpha_{min}$)when $\epsilon$ tends to zero. For a monofractal, this interval reduces to a point.

Again, the number $N_{\epsilon}(\alpha)$ of boxes of size $\epsilon$ that have a $\textit{H\"{o}lder}$ exponent between $\alpha$ and $\alpha +\delta\alpha$ obeys a power law as
\begin{equation}
N_{\epsilon}(\alpha) \propto \epsilon^{-f(\alpha)}
\label{f_alpha} 
\end{equation} 
where $f(\alpha)$ is a scaling exponent of the boxes with a common $\alpha$, called the singularity exponent. A plot of $f(\alpha)$ 
versus $\alpha$ is called the singularity spectrum. $f(\alpha)$ is the fractal dimension of the set of points that have the same 
singularity exponent $\alpha$. There can be several such interwoven fractal sets of points each with its particular value of 
$f(\alpha)$. Within each such set, the measure shows a particular scaling described by $\alpha$.

 Following (Chhabra et al., 1989), the functions $\alpha$ and $f(\alpha)$ can be determined 
by Legendre transformation as 
\begin{equation}
\alpha(q) = -\dfrac{\tau(q)}{q}\quad  \text{and}\quad  f(\alpha) = \alpha(q)q - \tau(q)
\label{legendre} 
\end{equation} 
Thus the singularity exponent defined by eq.(\ref{alpha}) becomes a decreasing function of $q$. Larger values of $q$ ($q>>1$) correspond 
to smaller exponents and therefore higher concentration of measure. Similarly smaller $q$ values correspond to higher exponents and lower concentration of measure. As $q$ varies, points ($\alpha_{q},f(\alpha_{q})$) define a parabolic curve that attains 
a maximum value $f(\alpha_0)$ at the point $\alpha_0$. $\alpha_0$ is the mean value of the singularity exponents and 
$f(\alpha_0)$ gives the fractal dimension of the support as obtained by the box-counting method.

Another equivalent description of the multifractal system is obtained from $D_q - q$ plot, where $D_q$, called the generalised 
dimension, corresponds to the scaling exponent for the $q^{th}$ moment of the measure.  It is defined by 
\begin{equation}
\displaystyle{{D_q}_{\epsilon \rightarrow 0}} \,=\,\frac{1}{1-q} \frac{log[\chi(q,\epsilon)]}{log(\epsilon)}
\label{D-q} 
\end{equation} 
For the particular case of $q=1$, eq.(\ref{D-q}) becomes indeterminate, and is estimated by l'H\^{o}pital's rule. 
$D_q$ is related to the mass exponent $\tau(q)$ by\
\begin{equation}
\tau(q) = (1-q)D_q 
\label{tau-D} 
\end{equation}
The generalised dimensions $D_q$ for $q=0$, $q=1$ and $q=2$ are known as the Capacity, the Information (Shannon entropy) and 
Correlation Dimensions respectively. Mathematically, the multifractals can be completely determined only by the entire multifractal spectrum. 
However a few characteristic functions may be used to describe the main properties of multifractals.

\section{ Multifractal analysis of sedimentary rocks}

\subsection{Simulated rock structure}

In the case of the simulated structure, a $256 \times 256\times 256$ cube was selected from below the deepest trough from the surface after the porosity had stabilized in an initial structure of $256 \times 256\times 3000$,
 and after the porosity had stabilized for a particular choice of $p$. This system was then 
covered with hypercubes of size $\epsilon = 2^k$ with $k$ ranging between $1$ to $16$.
The partition function $\chi(q,\epsilon)$, calculated according to eq.(\ref{chi}) for different values of box size $\epsilon$ 
and for different moment values $q$ was determined. A log-log plot of $\chi(q,\epsilon)$ versus $\epsilon$ when plotted, showed a deviation from linearity beyond a certain range of $\epsilon$. A power 
law scaling was observed only in the range $\epsilon = 2^1$ to $2^5$.  All calculations have been done within this range of 
$\epsilon$ for $q$ ranging from $-9$ to $+9$. The exponent $\tau(q)$ for each such $q$ and for every $p$ studied was noted.

The scaling properties observed in the partition function can be be characterized by determining if the scaling is simple 
as in monofractal, or multiple as in multifractal. Figs.(\ref{fig5}a and \ref{fig5}b) show the variation of $\tau(q)$ versus $q$ for  a low porosity $\phi=0.07$ corresponding to $p=0.9$, and a high porosity $\phi=0.42$ corresponding to $p=0.5$ respectively. The data points from simulation studies is shown as open circles in both the graphs. The plots for all the other $p$ values studied, lie 
within the limits set by these two plots. It is clear that the $\tau(q)$ functions which would have been straight lines for monofractals, deviate from linear behaviour. Moreover the slopes of the $\tau(q)$ for $q<0$ are quite different from those for $q>0$. This 
clearly indicates multiple scaling behaviour, i.e. the low density and high density regions of pores scale differently.

Subsequently, the generalised dimension $D_q$ were estimated in the range of $q$ values from $+9$ to $-9$. Figs.(\ref{fig6}a and \ref{fig6}b) 
show plots of $D_q$ versus $q$ for a low porosity corresponding 
to $p=0.9$ and a high porosity value corresponding to $p=0.5$ respectively. The data points from the 
simulation study are shown as open circles in the figure. For monofractals, all the $D_q$s would lie on the same horizontal line. In the case of the simulated sedimentary 
rock, the first three generalised dimensions are different for every $p$ (hence porosity), as shown in fig.(\ref{fig7}). This indicates that the structure is multifractal.  $Table I$ shows the results 
of the generalised dimensions for the first three moments calculated for the simulated structure at different porosities. It is apparent that at lower $p$ values,  between $0$ to $0.6$, the first three generalised dimensions have values very close to each other. The variation of porosity with $p$ is also very low. This indicates that the structure is quite homogeneous here.

The capacity dimension, $D_0$ provides information about how abundantly the measure,defined by eq.(\ref{measure}), is distributed over the scales of interest. 
Except for an abrupt increase at $p = 0.7$, the value of $D_0$ remains almost the same indicating that the same pore abundance is present at all the 
length scales studied. $D_0$ shows a maximum (fig.\ref{fig7})at $p=0.7$. In an earlier study (Sadhukhan et al., 2007b) on conductivity through connected pore space of sedimentary rocks, the effective conductivity of the simulated rock using $RBBDM$, showed a maximum for $p=0.7$ despite a maximum porosity at $p=0.5$. For structures generated by using the $RBBDM$, 
the maximum backbone mass of the connected cluster corresponded to $p=0.7$. The authors had established that it was the backbone mass of the connected cluster that was most effective for transport. The maximum value of $D_0$ at $p=0.7$ indicates that the capacity dimension is directly related to the pore distribution in the backbone of the connected cluster.

From $Table I$, it is seen that the entropy or information dimension $D_1$  decreases monotonically with increasing $p$. Lower 
$D_1$ indicates greater concentration of pores over a small size domain, i.e. greater clustering.  When $D_1$ is close to $0$, it will be be reflected as a sharp peak on a pore size distribution curve. Comparison of figs.(\ref{fig2} and \ref{fig3}) clearly indicate that there is greater clustering of pores with decreasing porosity. For the particular case of $p=0.7$ where conductivity through simulated structures using $RBBDM$ showed a maximum, it may be noted that the difference $D_0 - D_1$ is a minimum. 
Here the Capacity Dimension $D_0$ is a maximum while the Entropy Dimension $D_1$ is a minimum, and this has optimized connectivity. This is manifested in transport property having maximum values here.

The correlation function $D_2$ describes the uniformity of the measure (here pore cluster size) among different intervals.
Smaller $D_2$ values indicate long-range dependence, whereas higher values indicate domination of short range dependence.
From $Table I$ we see that $D_2$ shows a slight decrease at lower porosities  which is indicative of long range correlations appearing 
between pores. This is a manifestation of our growth algorithm for the sedimentary rocks. When the fraction of larger grains is small, elongated and isolated pore clusters are more prominent. Thus the pore sites show greater auto-correlation along these 
clusters. When porosity increases, though $D_2$ increases somewhat, it is not too significant as the pore clusters still retain 
their elongated appearance in spite of greater connectivity between the pores.

The $\alpha$ and $f(\alpha)$ values of the singularity spectrum were computed with the help of eq.(\ref{legendre}). The plot of 
$f(\alpha)$ versus $\alpha$ is shown in fig.(\ref{fig8}) for the different values of $p$ studied. The shape and symmetry parameters from the singularity curves is listed in $Table I$ for different values of $p$. The $\textit{H\"{o}lder}$ exponent $\alpha_0$ for each $p$ gives the average values of local mass distribution for  a given scale.  A greater value of $\alpha_0$ indicates a lower degree of mass concentration. This in turn would indicate that the pore distribution is highly 
heterogeneous and anisotropic with fluctuations in local densities. From $TableI$ it appears that $\alpha_0$ remains almost constant for different porosities of the simulated structures except for $p=0.7$, where it is a maximum.

Width of the $f(\alpha)$ spectra is defined as the difference between the $\alpha$ values of the most negative moment $q_{-}$ i.e. $\alpha_{max}$, and the most positive moment $q_+$ i.e. $\alpha_{min}$. The wider the spectrum, i.e. greater the difference between $(\alpha_{max} - \alpha_{min})$, 
the higher is the heterogeneity in the scaling indices of pore mass and vice versa. The largest $f(\alpha)$ obtained for $q=0$ 
corresponds to the capacity dimension $D_0$. Small $f(\alpha)$ values indicate rare events (extreme values of the PSD). Asymmetry
in the $f(\alpha)$ spectra indicate the dominance of higher or lower values of pore masses. If the width on the left, 
$\alpha_0 - \alpha_{min}$ is larger, this indicates the domination of large values in the PSD. From $Table I$ we see 
that $\alpha_0 - \alpha_{min}$ is larger than $\alpha_0 - \alpha_{max}$ for $p$ values between $0.1$ to $0.6$. This implies that 
there is greater dominance of pore masses here. On the other
hand, a large right width $\alpha_0 - \alpha_{max}$ would indicate the dominance of extremely small values in the PSD. For the 
porosities studied, it is clear from $Table I$ that this occurs for $p$ values between $0.7$ to $0.9$. In this 
region clustering of pore sites into elongated isolated channels leaving larger sections of structure pore free. 
Fig.(\ref{fig2}) illustrates this arrangement of pore clusters. The $f(\alpha)$ spectra is almost symmetric about $f(\alpha_0)$. 
This indicates that the rock structure has the most isotropic  pore distribution here. 
\subsection{Real rock structure}

\begin{figure}[!h]
\begin{center}
\includegraphics[scale=0.6]{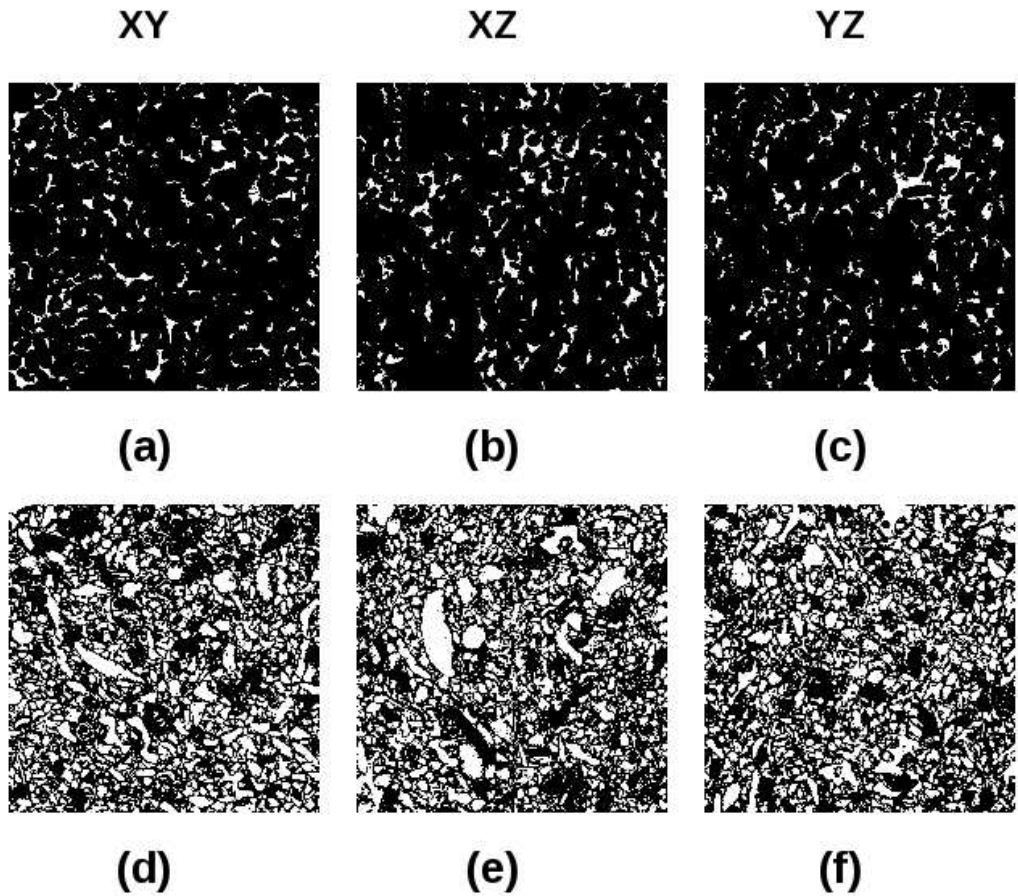}
\caption{Each section is a square of side $2.58\times 10^{-3}m$.(a) and (b) show sections of the real limestone cut in mutually perpendicular planes. These are perpendicular to bedding 
planes. (c) shows a section of the bedding plane. These are sections of oolitic limestone(pure calcite)from the Mondeville formation of Middle Jurassic age (Paris
Basin, France). (e) and (f) show sections of reefal carbonate obtained from Majorca Islands, Spain, perpendicular to bedding 
planes. (f) shows a section of the carbonate rock along the bedding plane.
 This rock structure looks more isotropic.} \label{fig4}
\end{center}
\end{figure}

\begin{figure}[!h]
\begin{center}
\includegraphics[scale=0.4]{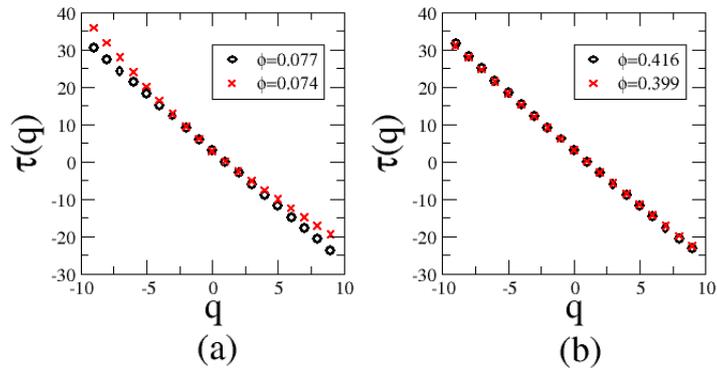}
\caption{(a)Variation of $\tau(q)$ versus $q$ for low porosity $\phi \simeq 0.07$. The open circles show data for simulated structure
while $X$ indicate data of real limestone sample. (b) Variation of $\tau(q)$ versus $q$ for high porosity $\phi \simeq 0.4$. The open circles show data for simulated structure
while $X$ indicate data of real carbonate sample.} \label{fig5}
\end{center}
\end{figure}

To compare our simulation results with real rock samples, X-ray tomography micrographs of $2-d$ sections of two real sedimentary rock samples obtained from an
oolitic limestone (pure calcite) from the Mondeville formation of Middle Jurassic age (Paris
Basin, France), and a reefal carbonate from the Majorca Islands, Spain, have been used. The limestone is composed of recrystallized oolite with a mean diameter of less
than a few hundred $\mu$m. Each pixel of the micrographs corresponds to $5.06$ micron.   
 Each section was converted to a binary file form such that $0$ corresponded to a pore site and $1$ 
 corresponded to a rock site. The binary file was then converted to a grey-scaled picture, as shown in 
 fig.(\ref{fig4}), 
 using MATLAB.
An array of $1000$ consecutive sections were put together precisely to reconstruct the binary file form of the 
real three dimensional rock structure. The real structure was $1000 \times 1000 \times 1000$ in size.  Figs.(\ref{fig4}a and \ref{fig4}b) show sections of the three dimensional limestone rock cut
  along the direction of assembly while fig.(\ref{fig4}c) shows a section of the same cut perpendicular to the direction of assembly.
  Similar sections were cut from the reefal carbonate and these are shown in figs.(\ref{fig4}d, \ref{fig4}e and \ref{fig4}f).
  The porosity of the oolitic limestone was determined from the reconstructed rock and found to be $0.073$ while the reefal carbonate was found to have a porosity of $0.399$. This high contrast in their porosity values is evident from the panels of 
 fig.(\ref{fig4}).

To compare the results of the $RBBDM$ for sedimentary rocks with real rocks, we have plotted the variation of 
$\tau(q)$ and  $D_q$  versus $q$ for the real rocks along with their closest matching porosity samples generated by the $RBBDM$.  
   The log-log plot of $\tau(q)$ versus $q$ for $q$ ranging from $-9$ to $+9$ for the low porous limestone is shown in 
   fig.(\ref{fig5}a) while the same plot for the high porous carbonate sample is shown in fig.(\ref{fig5}b).  The non-linear nature of the 
plots with two distinct slopes for positive and negative $q$ values clearly indicate that the real rock samples are also  
multifractal. It is clear that at high porosities, the real and the simulated rock give a very good match. The 
simulated rock at high porosities, fig.(\ref{fig2}), look more isotropic and start resembling real samples.
 For very low porosity 
like $0.073$, the $RBBDM$ does not yield a realistic rock sample. The long narrow pore clusters, fig.(\ref{fig3}a and \ref{fig3}b), an 
artefact of the generation rule, are responsible for a pronounced anisotropy which results in this mismatch. 

\begin{figure}[!h]
\begin{center}
\includegraphics[scale=0.65]{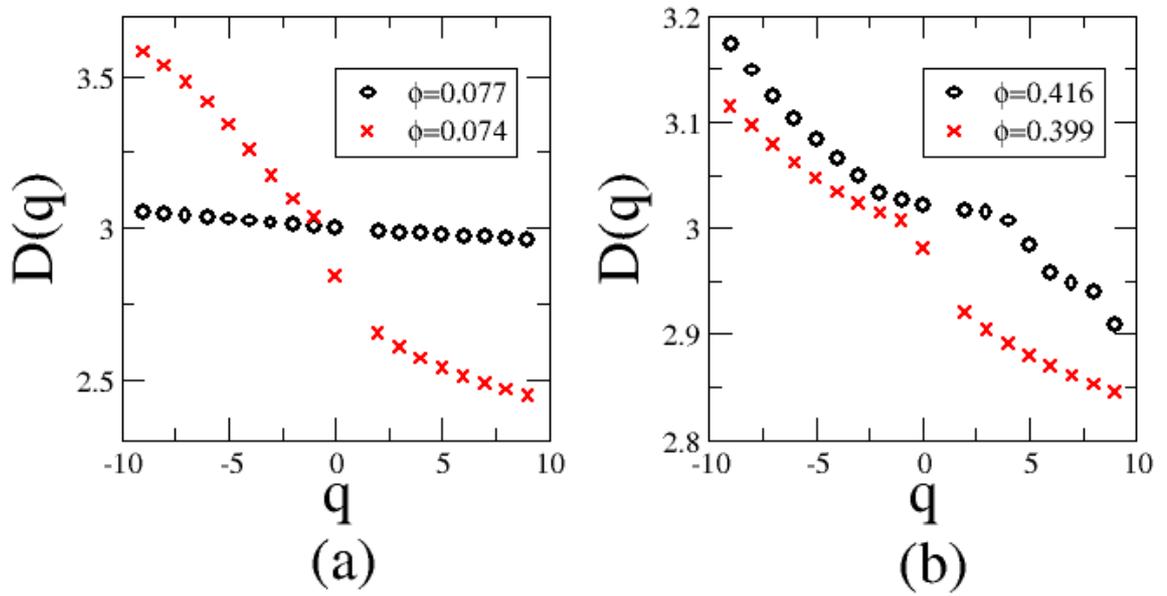}
\caption{(a)Plot of $D-q$ versus $q$ for low porosity corresponding 
to $\phi \simeq 0.07$. The open circles show data for simulated structure
while $X$ indicate data of real limestone sample. (b)$D-q$ versus $q$ for high porosity $\phi \simeq 0.4$. The open circles show data for simulated structure
while $X$ indicate data of real carbonate sample.} \label{fig6}
\end{center}
\end{figure}

\begin{figure}[!h]
\begin{center}
\includegraphics[scale=0.65]{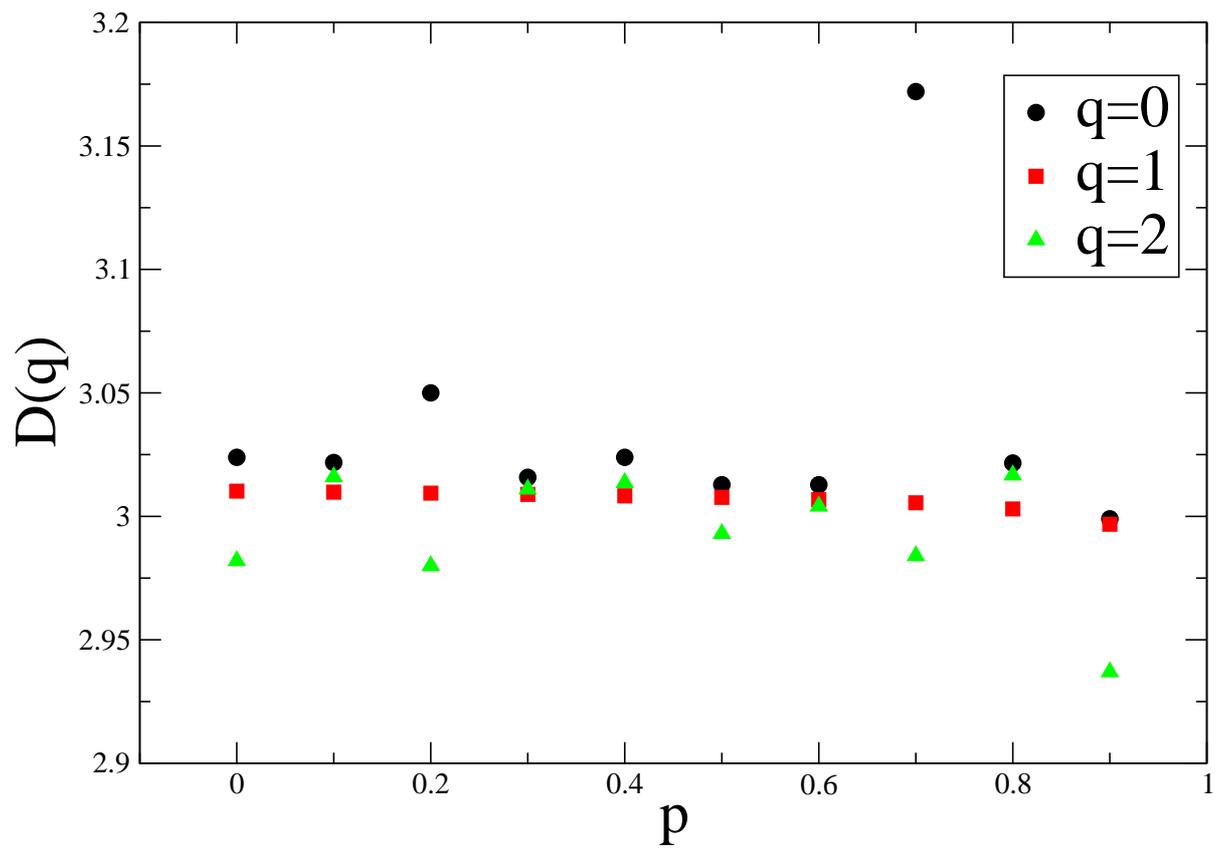}
\caption{Variation of the first three generalised dimension with porosity for simulated rock structure.}
 \label{fig7}
\end{center}
\end{figure}

\begin{figure}[!h]
\begin{center}
\includegraphics[scale=0.65]{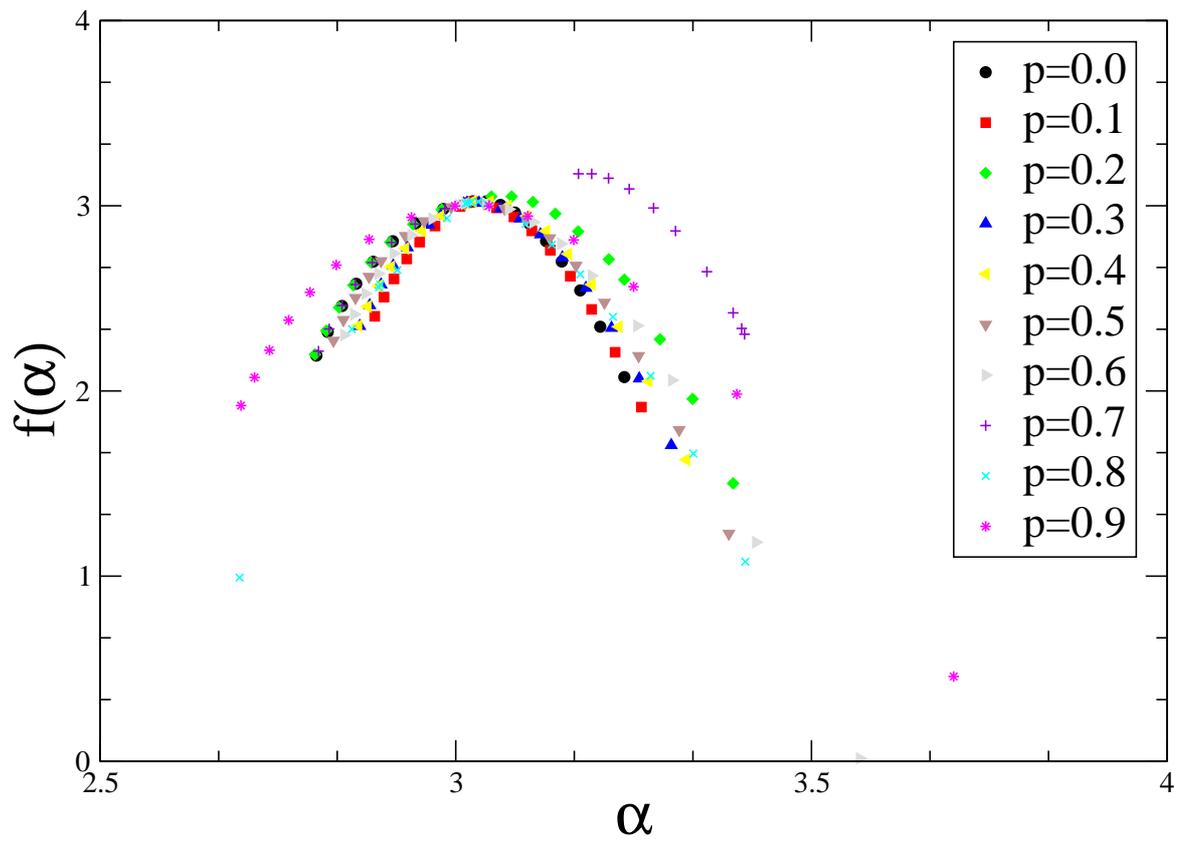}
\caption{Plot of 
$f(\alpha)$ versus $\alpha$ is for different porosities of simulated rock.
} \label{fig8}
\end{center}
\end{figure}

The plots of $D_q$ versus $q$ for the same samples over the same range of $q$ values are shown in figs.(\ref{fig6}a and \ref{fig6}b) 
 along with their corresponding matches from the $RBBDM$ structures.  Once again the similarity between the real and simulated rocks at high porosity values is clear.  Even for high porosity, fig.(\ref{fig6}b), the $D_q$ values at higher positive $q$ values show 
 a mismatch between real and simulation study. An examination of figs.(\ref{fig2} and \ref{fig4}d) reveal that pore cluster size and shape distribution is quite different even though the porosity values match. The 
generalised dimensions corresponding to the first three moments in each of the two real rocks,are enlisted in $Table II$ and 
$Table III$. It is clear that $D_0$, $ D_1$ and 
 $D_2$ are very different from each other in the case of the limestone sample 
showing clear multifractal nature.  The difference between the first three moments of the carbonate rock though 
less pronounced, is finite. Though both  the real rocks have a Capacity Dimension $D_0$ of almost the same value, the limestone has a lower value of $D_1$ in comparison to the carbonate. This indicates greater clustering of pores in the limestone sample.  One can 
expect that the limestone will be more efficient for fluid transport than the carbonate sample.  The smaller $D_2$ value of the 
limestone indicates that there is greater long range correlation between the pore clusters here than in the 
case of the carbonate sample.

\begin{figure}[!h]
\begin{center}
\includegraphics[scale=0.65]{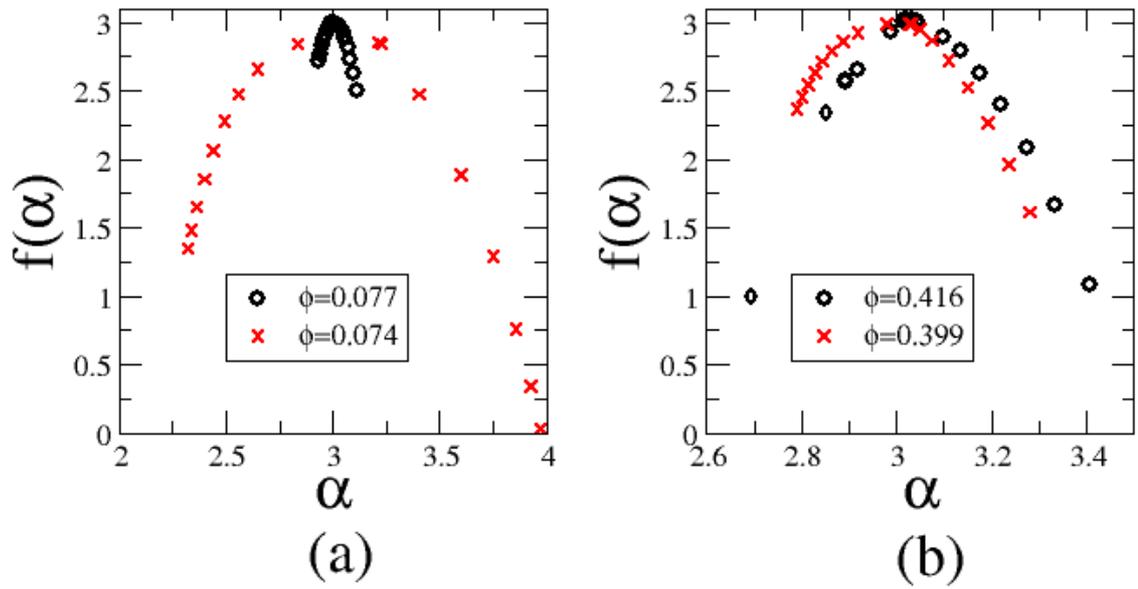}
\caption{(a)$f(\alpha)$ versus $\alpha$ for low porosity corresponding 
to $\phi \simeq 0.07$. The open circles show data for simulated structure
while $X$ indicate data of real limestone sample. (b)$f(\alpha)$ versus $\alpha$ for high porosity $\phi \simeq 0.4$. The open circles show data for simulated structure
while $X$ indicate data of real carbonate sample.   }
 \label{fig9}
\end{center}
\end{figure}

 The $f(\alpha)$ spectra of both the real rocks along with their corresponding simulated rock structures having similar 
 porosity, are shown in fig.(\ref{fig9}). The real and simulated rock show similar $f(\alpha)$ spectra at high 
 porosity. At very low porosity, the $RBBDM$ fails to create realistic sedimentary rocks. Both the real samples have a wider width 
 of their $f(\alpha)$ spectra indicating that there is a greater heterogeneity in the scaling indices 
 of their pore mass. With the $RBBDM$, this nature is observed as porosity increases. It is clear from the fig.(\ref{fig9}a)
  as also from $Table II$ 
that $\alpha_0 - \alpha_{max}$ is greater than  $\alpha_0 - \alpha_{min}$ in the limestone sample. This indicates that there is greater dominance of extremely small values in the PSD. Not only is the porosity of the oolitic limestone small, the pore clusters are small and sparse. This is also observed in figs.(\ref{fig4}a, \ref{fig4}b and \ref{fig4}c). $Table III$ indicates that 
in the carbonate rock, the difference between $\alpha_0 - \alpha_{max}$ and $\alpha_0 - \alpha_{min}$ is quite pronounced. The 
larger value of $\alpha_0 - \alpha_{max}$ shows a greater dominance of smaller values in the $PSD$. This dominance of 
smaller pore clusters is also seen from figs.(\ref{fig4}d, \ref{fig4}e and \ref{fig4}f). Large pore clusters are far and in between here.

\section{Conclusions}
Multifractal analysis on sedimentary rock structures simulated by using $RBBDM$ was done. The structures at different porosities, all showed multifractal 
characteristics. The complex heterogeneity of the pore size distribution has been quantified by the multifractal parameters. The Capacity Dimension gives a measure of the pore distribution in the backbone of the connected cluster. Fluid transport through such rocks maybe related to the multifractal parameters $D_0$ and $D_1$. A combination of 
higher $D_0$ and lower $D_1$ will result in more efficient transport properties.

Multifractal analysis performed on real sedimentary rock samples showed that these too, were multifractal in nature.  A comparison of the multifractal 
characters of both the real rocks studied, and their corresponding simulated structures with almost matching porosities, was done. The $RBBDM$ showed a very good match with the real sample at high porosity values for all the multifractal parameters.
At very low porosities however, even though both the simulated and real samples showed multifractal nature, the character match 
was not so good. At very low porosities the anisotropic nature of the simulated structure becomes more pronounced. A more suitable toppling rule may perhaps reduce this anisotropy somewhat. In the case of real rocks with low porosity, the pore 
distribution 
remains  more isotropic. 
We may conclude that the $RBBDM$ 
may be considered a good model for sedimentary rock generation especially at high porosity values as it shows similar geometric features as real rocks.

\section{Acknowledgement}
  This work is supported by Indo-French Centre For the promotion Of Advanced research (IFCPAR project no:4409-1). A. Giri is grateful to IFCPAR  for providing a research fellowship.

 \section{References} 
Chhabra,A.B., Meneveau,C., Jensen,R.V., Sreenivassen,K.R., 1989, Direct determination of the $f(\alpha)$ singularity spectrum and its application to fully developed turbulence,\textit{Phys.Rev.A} 40,5284-5294,$DOI:10.1103/PhysRevA.40.5284$.\\
Dathe, A., Tarquis, A.M., Perrier,E., 2006, Multifractal analysis of pore and solid phases in binary two-dimensional images of natural porous structures., \textit{Geoderma} 134,318-326,$doi:10.1016/j.geoderma.2006.03.024$.\\
Dutta Tapati, Tarafdar,S., 2003, Fractal pore structure of sedimentary rocks: Simulation by ballistic deposition, \textit{J. Geophysical Res.}, 108, NO. B2, 2062,$doi:10.1029/2001JB000523$.\\
Grau, J., Mendez,V., Tarquis,A.M., Diaz,M.C., Saa,A., 2006, Comparison of gliding box and box-counting methods in soil image analysis., \textit{Geoderma}, 134, 349-359,$doi:10.1016/j.geoderma.2006.03.009$.\\
Giri, A., Tarafdar,S., Gouze,P., Dutta,T., 2012a, Fractal pore structure of sedimentary rocks: Simulation in 2-d using a relaxed bidisperse ballistic deposition model, \textit{J.of Appl. Geophys.},87,40–45,$doi:10.1016/j.jappgeo.2012.09.002$. \\
Giri, A., Tarafdar,S., Gouze,P., Dutta,T., 2012b, Fractal geometry of sedimentary rocks: Simulation in 3D using a Relaxed Bidisperse Ballistic Deposition Model, Accepted for publication in \textit{Geophysical J. Int} on 2012 November 22 .\\
Manna.S.S., Dutta,T., Karmakar,R., Tarafdar,S., 2002, A percolation model for diagenesis, \textit{Int. J. Mod. Phys. C},13,319-331(2002),$DOI: 10.1142/S0129183102003176$.\\
Paz Ferriero,J., Wilson,M., Vidal Vazquez,E., 2009, Multifractal description of nitrogen adsorption isotherms.,\textit{Vadose Zone J.}, 8, 209-219 ,$doi:10.2136/vzj2008.0007 $\\
PettijohnF.J., 1984, Sedimentary Rocks, Harper \& Row Publishers Inc., U.S.A.\\ 
Rieu,M., Sposito,G., 1991, Fractal fragmentation, soil porosity and soil water properties.1.Theory, \textit{Soil Sci. Soc. Am. J.}, 67, 1361-1369,$doi:10.2136/sssaj2003.1361$.\\
Sadhukhan.S., Dutta,T., Tarafdar,S., 2007a, Simulation of diagenesis and permeability variation in two-dimensional rock structure., \textit{Geophys. J. Int.},169, 1366–1375,$DOI: 10.1111/j.1365-246X.2007.03426.x$.\\
Sadhukhan.S., Dutta,T., Tarafdar,S., 2007b, Pore structure and conductivity modelled by bidisperse ballistic deposition with relaxation., \textit{Modeling Simul. Mater. Sci. Eng.},15, 773-786,$doi:10.1088/0965-0393/15/7/005$.\\
Sadhukhan.S., Mal,D., Dutta,T., Tarafdar,S., 2008,Permeability variation with fracture dissolution: Role of diffusion vs. drift.,\textit{Physica A} ,387, 4541-4546,$doi:10.1016/j.physa.2008.03.026$. \\
Sadhukhan.S., Gouze,P., Dutta,T., 2012, Porosity and permeability changes in sedimentary rocks induced by injection
of reactive fluid: A simulation model.,\textit{J. Hydrol.},450,  134-139,$doi: 10.1016/j.jhydrol.2012.05.024$\\
Stauffer.D., Aharony,A., Introduction to percolation theory, $2^{nd}$ edition, 1994, Taylor and Francis, UK,$isbn:9780748402533$.\\ 
Tarafdar.S., Roy,S., 1998, A growth model for porous sedimentary rocks,\textit{Physica B} ,254, 28-36,$PII: S 0 9 2 1 - 4 5 2 6 ( 9 8 ) 0 0 4 3 1 - 1$.\\
Tarquis, A.M., Gimenez,G., Saa,A., Diaz,M.C., Gasco,J.M., 2003, Scaling and multiscaling of soil pore systems determined by image analysis., p19-34, \textit{In J. Pachepsky et al. (ed.)}, Scaling methods in soil physics., CRC Press, Boca raton, Fl,$DOI: 10.1201/9780203011065.ch2$.\\
Tarquis, A.M., Heck,R.J., Grau,S.B., Fabregat,,J., Sanchez,M.B., Anton,J.M., 2007, Influence of thresholding in mass and entropy dimension of $3-D$ soil images, \textit{Nonlin. Processes Geophys.}, 15,881-891,$doi:10.5194/npg-15-881-2008$.\\
Vidal Vazquez, E., Paz Ferreiro,J., Miranda,J.G.V.,  Paz Gonzalez,A., 2008, Multifractal analysis of pore size distributions as affected by simulated rainfall., \textit{Vadose Zone J.}, 7, 500-511,$doi:10.2136/vzj2007.0011$.\\


\begin{table}
\caption{ Multifractal parameters for simulated rock.}

\begin{tabular}{|c|c|c|c|c|c|c|c|c|c|}
\hline p & $\phi$ & $d_{0}$ & $d_{1}$ & $d_{2}$ & $\alpha_{0}$ & $\alpha_{min}$ & $\alpha_{max}$ & $\alpha_{0}-\alpha_{min}$ & $\alpha_{max}-\alpha_{0}$ \\ 
\hline 0.0000 & 0.4260 & 3.0239 & 3.0102 & 2.9820 & 3.0239 & 2.8040 & 3.2370 & 0.2199 & 0.2131 \\ 
\hline 0.1000 & 0.4342 & 3.0218 & 3.0098 & 3.0160 & 3.0218 & 2.8859 & 3.2609 & 0.1359 & 0.2391 \\ 
\hline 0.2000 & 0.4428 & 3.0500 & 3.0094 & 2.9800 & 3.0500 & 2.8020 & 3.3900 & 0.2480 & 0.3400 \\ 
\hline 0.3000 & 0.4498 & 3.0158 & 3.0089 & 3.0110 & 3.0158 & 2.8649 & 3.3029 & 0.1509 & 0.2871 \\ 
\hline 0.4000 & 0.4541 & 3.0239 & 3.0083 & 3.0136 & 3.0239 & 2.8630 & 3.3239 & 0.1609 & 0.3000 \\ 
\hline 0.5000 & 0.4557 & 3.0128 & 3.0077 & 2.9930 & 3.0128 & 2.8280 & 3.3840 & 0.1848 & 0.3712 \\ 
\hline 0.6000 & 0.4525 & 3.0128 & 3.0068 & 3.0041 & 3.0128 & 2.8430 & 3.5679 & 0.1698 & 0.5551 \\ 
\hline 0.7000 & 0.4410 & 3.1720 & 3.0055 & 2.9840 & 3.1720 & 2.8070 & 3.4059 & 0.3650 & 0.2339 \\ 
\hline 0.8000 & 0.4162 & 3.0216 & 3.0030 & 3.0167 & 3.0216 & 2.6959 & 3.4069 & 0.3257 & 0.3853 \\ 
\hline 0.9000 & 0.3518 & 2.9990 & 2.9968 & 2.9370 & 2.9990 & 2.6979 & 3.7000 & 0.3011 & 0.7010 \\ 
\hline 
\end{tabular}

\end{table}

\begin{table}
\caption{ Multifractal parameters for oolitic limestone (pure calcite) from the Mondeville formation of Middle Jurassic age (Paris
Basin, France).A system size $512^3$ was coarse grained to $64^3$.}
						
\begin{tabular}{|c|c|c|c|c|c|c|c|c|}
\hline $\phi$ & $d_{0}$ & $d_{1}$ & $d_{2}$ & $\alpha_{0}$ & $\alpha_{min}$ & $\alpha_{max}$ & $\alpha_{0}-\alpha_{min}$ & $\alpha_{max}-\alpha_{0}$ \\ 
\hline 0.077 & 2.9727 & 2.216 & 2.6942 & 2.9727 & 2.0889 & 5.1529 & 0.8838 & 2.1802 \\ 
\hline 
\end{tabular} 
\caption{ Multifractal parameters for reefal carbonate from Majorca island (spain). A system size $1000^3$ was evaluated.}
						
\begin{tabular}{|c|c|c|c|c|c|c|c|c|}
\hline $\phi$ & $d_{0}$ & $d_{1}$ & $d_{2}$ & $\alpha_{0}$ & $\alpha_{min}$ & $\alpha_{max}$ & $\alpha_{0}-\alpha_{min}$ & $\alpha_{max}-\alpha_{0}$ \\ 
\hline 0.399 & 2.989 & 2.970 & 2.9603 & 2.989 & 2.898 & 3.329 & 0.091 & 0.34 \\ 
\hline 
\end{tabular}

\end{table}

\end{document}